\documentclass[iop]{emulateapj}

\def\mdot{\hbox{$\dot {\it M}$}}
\def\micron{$\mu$m}
\def\microns{$\mu$m}

\def\msun{M_{\sun}}

\def\h2{H$_2$}

\newcommand\msunyr{\rm {\it M}_{\odot}\,yr^{-1}}
\newcommand\be{\begin{equation}}
\newcommand\en{\end{equation}}

\usepackage{graphicx}
\usepackage[caption=false]{subfig}

\newcounter{column_number}
\begin{document}

\shortauthors{Espaillat et al.}
\shorttitle{The Star-Disk-Jet Connection in GM Aur}

\title{Revealing the Star-Disk-Jet Connection in GM Aur Using Multiwavelength Variability}

\author{
C. C. Espaillat\altaffilmark{1},
E. Mac{\'i}as\altaffilmark{1},
J. Hern{\'a}ndez\altaffilmark{2}, \& 
C. Robinson\altaffilmark{1}  
}

\altaffiltext{1}{Department of Astronomy \& Institute for Astrophysical Research, Boston University, 725 Commonwealth Avenue, Boston, MA 02215, USA; cce@bu.edu, emacias@bu.edu, connorr@bu.edu}
\altaffiltext{2}{Instituto de Astronom{\'i}a, Universidad Aut{\'o}noma Nacional de M{\'e}xico, Campus Ensenada; MX; hernandj@astrosen.unam.mx}

\begin{abstract}

Here we analyze the first simultaneous X-ray, ultraviolet, optical, infrared, and centimeter observations of a T Tauri star (TTS). We present three epochs of simultaneous {\it Spitzer} and VLA data of GM Aur separated by $\sim$1~wk.  These data are compared to previously published {\it HST} and {\it Chandra} observations from which mass accretion rates ($\mdot$) and X-ray luminosities, respectively, were measured. The mid-infrared emission increases along with $\mdot$, and we conclude that this is due to an increase in the mass in the inner disk. The cm emission, which probes the jet, also appears to increase as $\mdot$ increases, and the changes in the cm flux are consistent with the variability in $\mdot$ assuming the mass-loss rate is $\sim10\%~\mdot$. The 3~cm emission morphology also appears changed compared with observations taken three years previously, suggesting that for the first time, we may be tracking changes in the jet morphology of a TTS. The X-ray luminosity is constant throughout the three epochs, ruling out variable high-energy stellar radiation as the cause for the increases in the mid-infrared or cm emission. Tying together the multiwavelength variability observed, we conclude that an increase in the surface density in the inner disk resulted in more mass loading onto the star and therefore a higher $\mdot$, which led to a higher mass-loss rate in the jet. These results stress the importance of coordinated multiwavelength work to better understand the star-disk-jet connection.

\end{abstract}

\keywords{accretion disks, stars: circumstellar matter, 
planetary systems: protoplanetary disks, 
stars: formation, 
stars: pre-main sequence}

\section{Introduction} \label{intro}

Low-mass pre-main-sequence stars (i.e., T Tauri stars; TTS) are highly energetic, and many are surrounded by circumstellar disks displaying the typical signatures of mass accretion onto the star and mass ejection via jets  \citep[e.g.,][]{frank14,hartmann16}. 
High-energy radiation is mainly traced via X-ray observations (since we cannot observe the extreme ultraviolet; EUV) which predominantly originates from the stellar corona \citep[][]{feigelson02,brickhouse10}. The inner dust edge of the disk emits brightly in the mid-infrared \citep[MIR; e.g.,][]{muzerolle03, espaillat10, tannirkulam07, mcclure13, millan16}. 
Accreting TTS (i.e., classical TTS; CTTS) have strong magnetic fields \citep[][]{donati09,johnskrull13} that truncate the inner disk and lead to accretion of material onto the star \citep{hartmann16}.  The most direct measurement of the 
mass accretion rate, $\mdot$, comes from extracting the excess near-UV (NUV) continuum emission due to accretion 
\citep[e.g.,][]{calvet98,ingleby11b}. 
CTTS also show signs of mass ejection in the form of jets \citep[e.g.,][]{frank14} 
with mass-loss rates of $\sim$10$\%$~$\mdot$ \citep{hartigan95, natta14}.  

Variability is a trademark of TTS. Variability is well known in \mdot \citep[e.g.,][Robinson \& Espaillat 2019; RE19]{venuti14,cody18,siwak18} and X-ray emission \citep[e.g.,][]{preibisch05,argiroffi11,flaccomio12,principe14,guarcello17}.  In the optical and MIR, the most common type of variability is from ``dippers,'' objects that have fading events lasting one to five days \citep{cody14}. 
Centimeter flux variability has been seen on timescales of 30 min to 15 yr in young stellar objects (YSOs), likely due to changes in the stellar magnetosphere and wind \citep{liu14}. 

\begin{deluxetable*}{ccccccc}[ht!]    
\tablecaption{{\it Spitzer} Observations of GM Aur \label{tab:spitzer}}
\tablehead{
\colhead{Epoch} & \colhead{Date} & \colhead{Start} & \colhead{End}  & \colhead{AOR ID} &  \colhead{[3.6]} & \colhead{[4.5]} \\
\colhead{} & \colhead{(UT)} & \colhead{Time (UT)} & \colhead{Time (UT)} & \colhead{} & \colhead{(mag)} & \colhead{(mag)}}
\startdata
4 & 2016-01-05 & 20:12:34 & 23:07:45 & 58455040	  & 8.11$\pm$0.01 & 7.99$\pm$0.01 \\
5 & 2016-01-09 & 13:35:20 & 16:31:35 & 58454784  & 8.03$\pm$0.01 & 7.90$\pm$0.01 \\
6 & 2018-01-04 & 05:22:25 & 08:17:04 & 64918272  & 8.05$\pm$0.01 & 7.95$\pm$0.01 \\
7 & 2018-01-11 & 04:47:01 & 07:42:56 & 64918528  & 7.88$\pm$0.01 & 7.77$\pm$0.01 \\
8 & 2018-01-19 & 03:42:21 & 06:39:31 & 64918784  & 8.11$\pm$0.01 & 8.00$\pm$0.01 
\enddata
\tablecomments{Here we adopt the epoch labels used in RE19 and \citet{espaillat19} which analyzed eight epochs of GM Aur.}
\end{deluxetable*}

\begin{deluxetable*}{ccccccc}[ht!]     
\tablecaption{VLA Observations of GM Aur \label{tab:vla}}
\tablehead{
\colhead{Epoch} & \colhead{Date} & \colhead{Start} & \colhead{End}  & \colhead{$F_{\rm 3 cm}$\tablenotemark{a}} & \colhead{rms} & \colhead{Ph. Cal.} \\
\colhead{} & \colhead{(UT)} & \colhead{Time (UT)} & \colhead{Time (UT)} & \colhead{($\mu$Jy)} & \colhead{($\mu$Jy beam$^{-1}$)} & \colhead{$F_{\rm 3 cm}$ (mJy)}}
\startdata
6 & 2018-01-04 & 06:04:42 & 09:16:05 & $80.6\pm3.7$ & 2.6 & $450\pm4$ \\
7 & 2018-01-11 & 04:57:02 & 08:08:26 & $95.2\pm5.4$ & 3.3 & $477\pm5$ \\
8 & 2018-01-19 & 04:09:59 & 07:21:23 & $92.7\pm4.4$ & 3.2 & $469\pm6$ 
\enddata
\tablenotetext{a}{The error listed here does not include the expected 5\% absolute flux uncertainty of the VLA at 3 cm, but it is included in Figure~\ref{fig:vla}.}
\end{deluxetable*}

Few studies simultaneously link the multiwavelength variability seen in TTS.  Here we present results from a program that simultaneously observed the X-ray, ultraviolet, optical, IR, and cm wavelengths of the TTS GM Aur using {\it Chandra}, {\it HST}, {\it Spitzer}, and the VLA. 
GM Aur was chosen for this study since it is one of the best characterized disks at cm wavelengths and it is the only TTS surrounded by a transitional disk \citep[i.e., an object with a large disk hole,][]{espaillat14} with a resolved jet \citep{macias16}.  In addition, it displays multiwavelength variability \citep[][RE19]{espaillat11, ingleby15, espaillat19}. 
The {\it HST} far-UV (FUV), NUV, optical, and NIR data were first presented by RE19 in a study of accretion variability.  The {\it Chandra} data were first presented by \citet{espaillat19} in a study of the FUV and X-ray connection.  Here we present the {\it Spitzer} and VLA data of GM Aur.  We focus mainly on analyzing the simultaneous {\it Chandra}, {\it HST}, {\it Spitzer}, and VLA observations which were taken over three epochs separated by about 1 wk. However, we also present two additional epochs of {\it Spitzer} data that do not have coordinated VLA data but do have simultaneous {\it HST} data from RE19 and partially simultaneous {\it Swift} data from \citet{espaillat19}.  

In Section 2, we present the {\it Spitzer} and VLA data and discuss the timing of the overall multiwavelength dataset, and in Section 3, we search for trends.
In Section 4, we discuss the connection between disk variability probed by {\it Spitzer} and stellar variability probed with {\it HST}, {\it Swift}, and {\it Chandra}.
We also consider jet variability seen with the {\it VLA} in light of the {\it HST}, {\it Chandra}, and {\it Spitzer} observations.
We present conclusions in Section 5.

\begin{deluxetable*}{cccccccc}[t]       
\tabletypesize{\scriptsize}
\tablewidth{0pt}
\tablecaption{{\it HST}, {\it Swift}, and {\it Chandra} Observations of GM Aur and Stellar Properties \label{tab:properties}}
\tablehead{
\colhead{Epoch} & \colhead{Telescope/} & \colhead{Identification} & \colhead{Date} & \colhead{Start Time} & \colhead{End Time} & \colhead{\mdot} & \colhead{L$_{X}$} \\
\colhead{} 
&\colhead{Instrument} &\colhead{ No.} &\colhead{(UT)} &\colhead{(UT)} &\colhead{(UT)} & \colhead{(10$^{-8}$$\msunyr$)} &  \colhead{($10^{30} {\rm erg s}^{-1}$)}
}
\startdata
4  	& {\it HST}$/$STIS 	&	14048 & 2016-01-05 & 20:38:03 & 23:00:42 & $1.021^{+0.009}_{-0.009}$ &
 --   \\

4  	& {\it Swift}$/$XRT	& 00034249002 & 2016-01-05 & 20:43:02 & 22:58:00 & -- &
 $3.4^{+0.9}_{-0.6}$   \\

4  	& {\it Swift}$/$XRT	& 00034249003 & 2016-01-06 & 00:00:00 & 14:42:00 & -- &
 $3.4^{+0.9}_{-0.6}$  \\

\hline
5  	& {\it Swift}$/$XRT	& 00034249004 & 2016-01-09 & 09:13:26 & 16:13:53  & -- & $17.1^{+1.2}_{-0.9}$  \\

5  	& {\it HST}$/$STIS & 14048 & 2016-01-09 & 13:40:55 & 16:02:33 & $0.768^{+0.008}_{-0.008}$ & --   \\

\hline
6  	& {\it Chandra}$/$ACIS & 20614 & 2018-01-04 & 05:49:29 & 09:38:50 & -- & $4.4^{+0.4}_{-0.5}$ \\

6  & {\it HST}$/$STIS & 15165 & 2018-01-04 & 06:10:54 & 08:34:18 & $0.564^{+0.007}_{-0.007}$ & --   \\

\hline
7  	& {\it Chandra}$/$ACIS & 20615 & 2018-01-11 & 04:45:07 & 08:32:11 & -- &  $4.1^{+0.4}_{-0.6}$   \\

7  	& {\it HST}$/$STIS 		& 15165 & 2018-01-11 & 05:03:19 & 07:26:40 & $1.961^{+0.012}_{-0.012}$ & --   \\

\hline
8  	& {\it Chandra}$/$ACIS 	& 20616     & 2018-01-19 & 03:18:12 & 07:03:10 & -- & $4.7^{+0.5}_{-0.6}$    \\

8  	& {\it HST}$/$STIS 		& 15165 & 2018-01-19 & 03:43:47 & 06:07:09 & $0.979^{+0.009}_{-0.009}$ & -- 
\enddata
\tablecomments{\mdot is from RE19 and L$_{X}$ is from  Espaillat et al. (2019).
Identification numbers for {\it HST} and {\it Chandra}$/${\it Swift} correspond to the Proposal ID and Observation ID, respectively.}
\label{tab:properties}
\end{deluxetable*}

\section{Observations and Data Reduction} \label{redux}

Here we report new {\it Spitzer} and VLA data.  These data were taken simultaneously with {\it HST} and {\it Chandra} or {\it Swift} data and were first reported in RE19 and  \citet{espaillat19}, respectively. In order to facilitate comparison between the works, we adopt the same epoch labels; RE19 and \citet{espaillat19} analyzed eight epochs of GM~Aur.  Here we present five epochs of {\it Spitzer} data taken in Epoch (E) 4 through E8 and three epochs of VLA data taken in E6 to E8.  The main focus of this paper is on the three epochs of {\it Chandra-HST-Spitzer} data taken in E6 to E8.  However, we also discuss the two epochs of {\it Swift-HST-Spitzer} data taken in E4 and E5.  We compare the timing of the datasets in  Section~\ref{sec:simultaneity}.

\subsection{{\it Spitzer} IRAC}

{\it Spitzer} IRAC was used to observe GM~Aur in the [3.6] and [4.5] channels twice in Program 11071 (PI: Espaillat) and three times in Program 13227 (PI: Espaillat). Details of the observations are listed in Table~\ref{tab:spitzer}.  

To avoid saturation, we used the fixed subarray readout mode with a time sampling of 0.1~s, resulting in 64 Basic Calibrated Data (BCD) frames.
We employed
a standard dither pattern with a four-position Gaussian to remove cosmic ray hits, bad pixels, latent images and pixel-to-pixel uncertainties. We developed an IDL script to combine the BCD frames using a median algorithm, avoiding the 54th frame, which has a skydark subtraction issue (IRAC Handbook\footnote{https://irsa.ipac.caltech.edu/data/SPITZER/docs/irac/ iracinstrumenthandbook/}). Using IRAF \texttt{daofind}, we performed aperture photometry on the combined frames with an aperture radius of 3 pixels and a sky annulus from 3 to 7 pixels. Finally, aperture correction was applied. Mean magnitudes for each epoch are listed in Table~\ref{tab:spitzer}. There were no significant departures outside the uncertainties during the observations. 

\subsection{VLA}

VLA continuum observations at 3 cm (X band) were taken on three visits as part of Project SP0708 (PI: Espaillat; Table~\ref{tab:vla}). Each observation featured $\sim155$ min of on-source time. The VLA was in its B configuration, and all antennas remained in the same position for the three visits. The observations were taken within 2~hr, ensuring that the differences in the {\it u,v} coverage were small. The quasar 3C147 was used as the flux and bandpass calibrator, while J0438+3004 was the phase calibrator.

The data were calibrated using the VLA pipeline in CASA (version 5.1.0). After flagging data with irregular phases or amplitudes, we obtained cleaned images using \texttt{tclean} in CASA (version 5.3.0). The \texttt{mtmfs} algorithm was used with $nterms=2$ (straight spectrum) and pointlike components (scales = 0). The images were then convolved to an angular resolution of $1\rlap.''5\times1\rlap.''5$ to ensure a proper comparison between epochs. Integrated flux densities were measured by fitting a Gaussian to the convolved images (Table \ref{tab:vla}).  

\subsection{Simultaneity of the Observations} \label{sec:simultaneity}

Observation times of the {\it Spitzer} and VLA data can be found in Tables~\ref{tab:spitzer} and~\ref{tab:vla}, respectively.  Times for data presented in RE19 and \citet{espaillat19} can be found in Table~\ref{tab:properties}.

For E4 and E5, the {\it Spitzer} and {\it HST} data are essentially simultaneous. The {\it Spitzer} and {\it HST} data were taken along with {\it Swift} data, which are partially simultaneous. 
The {\it Swift} data were performed in two observations. For E4, the first {\it Swift} observation (00034249002) was taken simultaneously with the {\it HST} and {\it Spitzer} observations.  The second set of {\it Swift} data for E4 (00034249003) was taken within 16~hrs after the {\it HST} and {\it Spitzer} observations.   For E5, roughly 40\% of the {\it Swift} observations were simultaneous with {\it HST} and {\it Spitzer} and the rest of the {\it Swift} data were taken less than $\sim$4.5~hrs before the start of the {\it HST} and {\it Spitzer} observations.
For E6, E7, and E8, the {\it Spitzer}, {\it HST}, {\it Chandra}, and VLA data are essentially simultaneous.

\begin{figure}     
\plotone{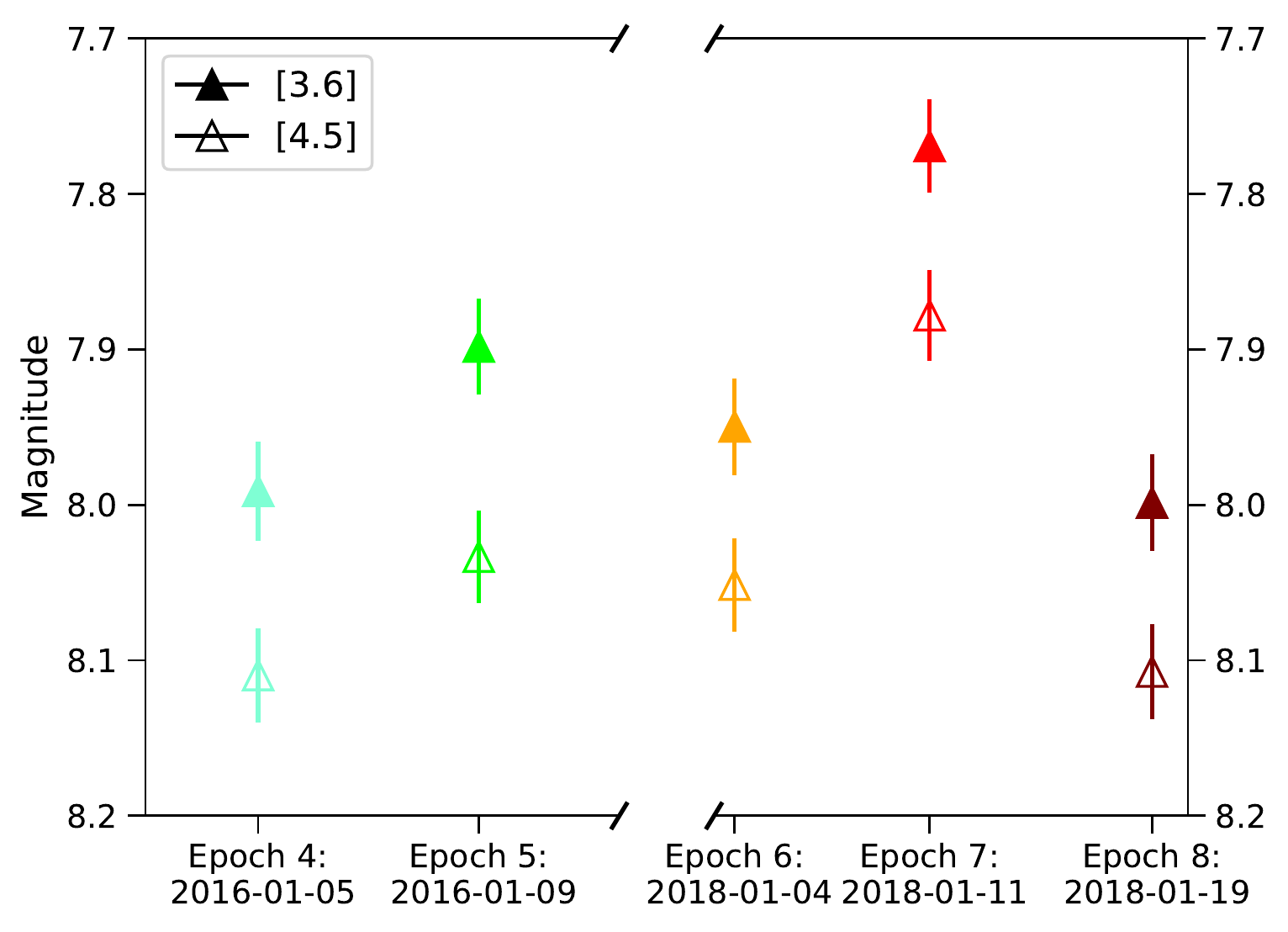}
\caption[]{
{\it Spitzer} IRAC [3.6] and [4.5] mean magnitudes of GM~Aur taken in E4 to E8.  There is a significant increase of emission in both {\it Spitzer} bands in E7.    
}
\label{fig:spitzer}
\end{figure} 

\section{Analysis and Results}

\subsection{Infrared Continuum Emission}

We dereddened the MIR data with the \citet{mathis90} extinction law using an R$_{V}$ of 3.1 and A$_{V}=0.6$ from \citet{manara14}. The E4, E5, E6, and E8 magnitudes are roughly consistent with each other and those seen on 2004 Feb 14 \citep[{[3.6]: $8.04\pm0.02$, [4.5]: $7.88\pm0.03$};][]{luhman10}.  The magnitude and range of the MIR observations are not atypical for GM~Aur aside from E7. In E7, the {\it Spitzer} IRAC [3.6] and [4.5] magnitudes are higher (Table~\ref{tab:spitzer}, Figure~\ref{fig:spitzer}), corresponding to a factor of $\sim1.2$ flux increase flux between E6 and E7. 

MIR emission shortward of $\sim$20~{\micron} has been attributed to optically thin dust within $\sim$1~au of GM~Aur \citep{calvet05,espaillat10}. \citet{espaillat11} found that GM~Aur's {\it Spitzer} spectra varied by about 10$\%$ and attributed this to variability in the mass of dust in the inner disk.  Using NASA IRTF SpeX spectra covering 1--5 \microns\, \citet{ingleby15} found that the emission varied and also attributed this to dust-mass changes.  The {\it Spitzer} spectra were not coordinated with {\it \mdot} indicators, but the SpeX data were taken within a day of {\it HST} UV data.  With three epochs of {\it HST} data, \citet{ingleby15} found that $\mdot$ and the IR emission are roughly the same between their first two epochs (taken about 1 wk apart), but both $\mdot$ and the IR emission decreased in the third epoch (taken about four months later). Their first and second epoch had $1.9\times10^{-12}$ $\msun$ of dust in the inner disk and the third epoch had $1.4\times10^{-12}$ $\msun$ of dust in the inner disk. 

\begin{figure*}                                    
\begin{minipage}[b]{.49\textwidth}
\includegraphics[scale=0.3]{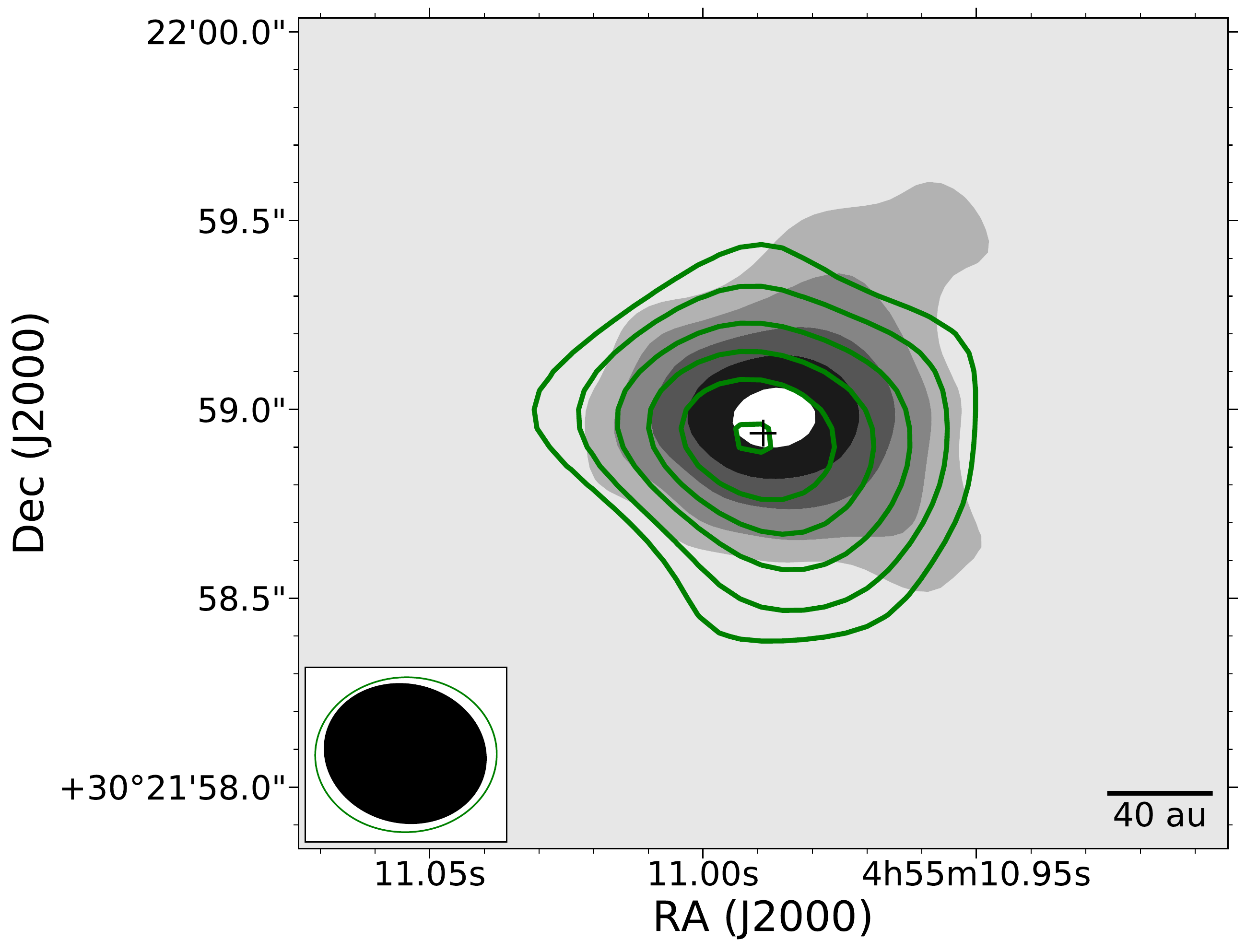}
\end{minipage}\hfill
\begin{minipage}[b]{.49\textwidth}
\includegraphics[scale=0.55]{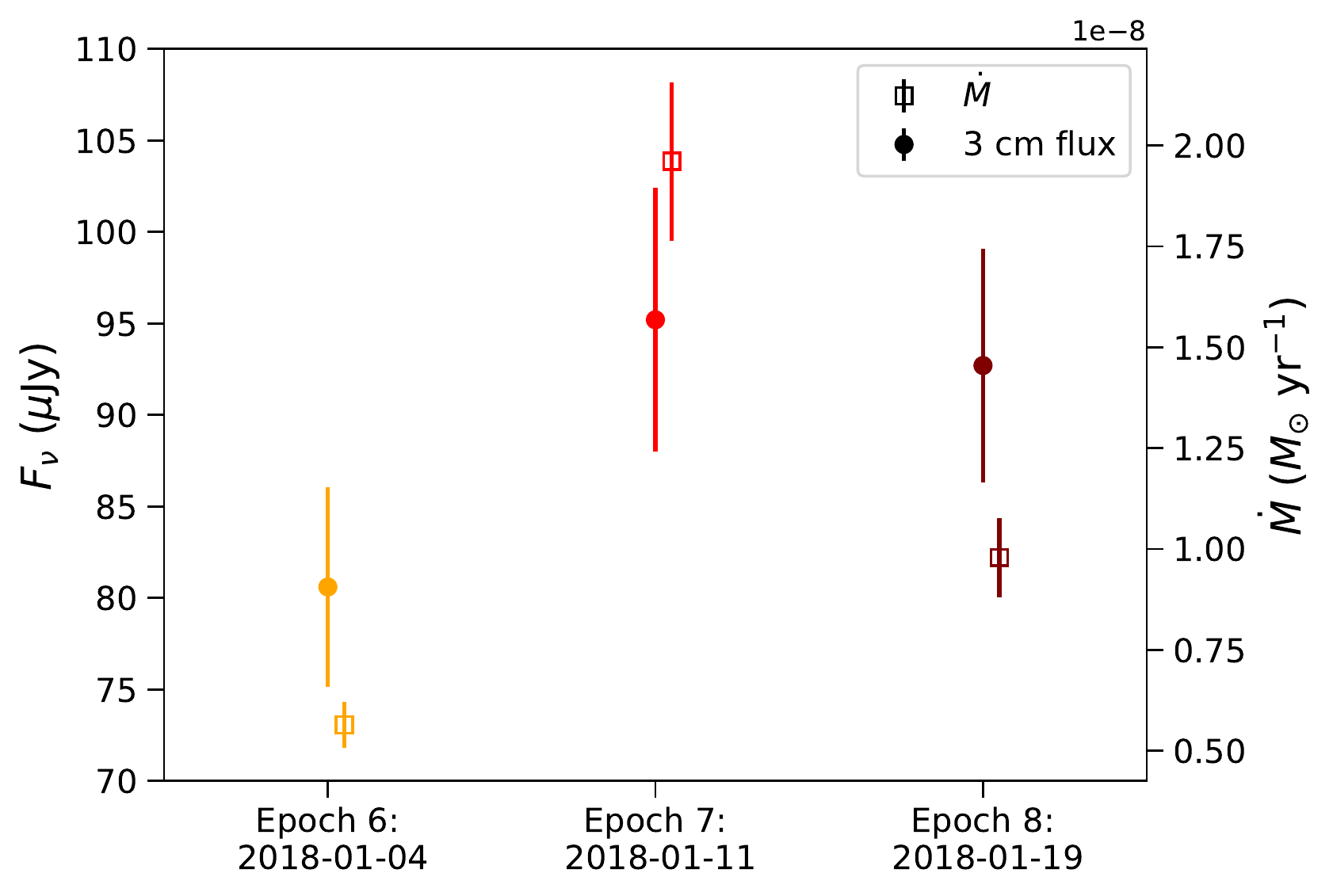}
\end{minipage}
\caption[]{{\it Left:} Superposition of 3 cm VLA observations of GM Aur from 2015 \citep[grayscale filled contours;][]{macias16} and E6 to E8 in 2018 (green contours). The contours indicate 3, 5, 8, 11, 14, and 17 times the rms of the maps (2.2 $\mu$Jy beam$^{-1}$ for 2015 and 1.8 $\mu$Jy beam$^{-1}$ for 2018). The beam sizes are shown in the lower left.
{\it Right:} \mdot in E6 to E8 compared to the 3 cm continuum flux.  The uncertainties in \mdot are about 10$\%$, assuming a visual extinction correction error of 0.5 from RE19.  The uncertainties in the VLA fluxes include the rms uncertainty (Table~\ref{tab:vla}) and the 5$\%$ absolute flux calibration uncertainty.
}
\label{fig:vla}
\end{figure*} 

Here we have only two MIR magnitudes, which is not ideal for constraining the optically thin dust model. However, we can estimate the amount of dust in the inner disk in E4 through E8 by scaling with the results of \citet{ingleby15} under the assumption that only the dust mass changes \citep{espaillat11,ingleby15}. The SpeX spectra cover 3.6~{\microns}, but not 4.5~{\microns} where atmospheric telluric lines are strong.  Convolving the SpeX spectra with the IRAC [3.6] bandpass, we determine that the SpeX flux in the first epoch of \citet{ingleby15} corresponds to an IRAC [3.6] magnitude of about 8.1, which is roughly consistent with our measurements in E4, E5, E6, and E8.  Therefore, we assume that the mass of dust in the inner disk in these four epochs was about the same as that seen in the first epoch of the \citet{ingleby15} observations. For E7, if we scale the optically thin model from the flux seen in the first epoch of the \citet{ingleby15} observations to the flux seen in E7, we estimate a dust mass of $\sim4.7\times10^{-12}$ $\msun$, corresponding to a factor of $\sim2.5$ change. 

\subsection{Centimeter Continuum Emission}

In Figure~\ref{fig:vla} (left), we compare the spatial brightness distribution of our 2018 data to 2015 data from \citet{macias16}.  We combine the three epochs of 2018 VLA data in this figure to achieve a high enough sensitivity to use robust weighting (r = \hbox{--0.5}) and a similar angular resolution to the 2015 data. There is a seeming change in the morphology of the jet emission between 2015 to 2018. The upper right emission that \citet{macias16} attributed to a jet appears to be present only in the 2015 data. The resolution and sensitivity of both maps are similar, suggesting that the difference in morphology is real. However, our resolution is limited, so higher spatial resolution observations are needed to confirm this.  

In Figure~\ref{fig:vla} (right), we plot the VLA fluxes of GM Aur from E6 to E8 (Table~\ref{tab:vla}) as well as $\mdot$ (Table~\ref{tab:properties}).  RE19 reported an accretion burst in E7.  We see a possible increase in the 3 cm flux in E7 as well. While the VLA uncertainties between E6 and E7 do not overlap, they are still close, and so we consider this speculatively moving forward.  

\section{Discussion} \label{sec:discussion}

\subsection{Variability in the Inner Disk} 

Here we see a large increase in the MIR emission of GM~Aur as $\mdot$ increases.  In E7, GM Aur was caught during a burst of accretion:  $\mdot$ in E6 was $\sim0.6\times10^{-8}$ $\msunyr$; seven days later in E7, it was $\sim2\times10^{-8}$ $\msunyr$; then it dropped down to $\sim1\times10^{-8}$ $\msunyr$ eight days later (Table~\ref{tab:properties}). We find that the dust mass increases by a factor of about 2.5, roughly consistent with the factor of 3.5 change in $\mdot$ between E6 and E7.  This supports our contention that the surface density changes in the inner disk affect the surface density at the star, consistent with work by \citet{ingleby15}.  

Further support for changes in the surface density are found by measuring the \h2 bump emission, which traces gas near the star. \citet{ingleby15} found that the luminosity of the \h2 bump emission decreased as $\mdot$ and the IR emission decreased.  \citet{espaillat19} measured the \h2 bump luminosity and found that it trends with $\mdot$ in a sample of 7 objects. In particular, it increases by a factor of $\sim2$ between E6 and E7, roughly consistent with the change in $\mdot$ and dust mass, supportive of surface density changes in the inner disk affecting the gas in the innermost disk and hence $\mdot$.

We can exclude that changes in MIR emission are due to an increase of irradiation from the accretion shock because the accretion luminosity is still much lower than the stellar luminosity (RE19). \citet{flaherty12} also found that MIR variability cannot be due to changes in the accretion luminosity. Increases in MIR emission could be due to an increase in X-ray heating \citep{glassgold04} or ionization \citep{fraschetti18}. Previous coordinated {\it Chandra} and IRAC variability studies of YSOs have not found a correlation between MIR and X-ray emission \citep{flaherty14, flaherty16b}. 
Our observations were simultaneous, and we also do not find a correlation between MIR and X-ray emission. Notably, E5 and E6 have similar MIR emission even though the X-ray luminosity changed significantly (Table~\ref{tab:properties}).  

We conclude that the emitting gas and dust originate in the same location in the disk and, therefore, that both gas and dust exist at the magnetospheric truncation radius where material accretes onto the star \citep[e.g.,][]{ingleby15}, contrary to the simple picture of magnetospheric accretion in CTTS. The need for dust near the co-rotation radius is also supported by larger variability studies, which find that the ``dippers'' must be explained by dust obscuration \citep{cody14,stauffer15}. Given that these innermost regions will not be resolved in the near future, time-domain studies are the best way to probe this region.

\subsection{Variability in the Jet}

GM Aur appears to have variability in the morphology and flux of the jet emission.  It is common to see changes in the jet morphology in protostars \citep[e.g.,][]{curiel06}. However, these would be the first detected changes in the jet emission of a TTS. More frequent monitoring over a longer time range is necessary to confirm these results and to analyze the time evolution of the 3 cm emission to see how it correlates with accretion variability.  Below, we speculate on the implications of these observations.

GM Aur was observed previously with the VLA at 0.7, 3, and 5~cm with high angular resolution in 2015 \citep{macias16}. Our 2018 VLA 3~cm observations do not show the tripolar morphology resolved by \citet[][Figure~\ref{fig:vla}, left]{macias16}. This change in the morphology of the jet emission could be due to the change in brightness of ejected material that was detectable in 2015 but not in 2018. If confirmed, monitoring of the radio jet could estimate the ejection velocities in the jet and the variability in ejection of material. 

There also seems to be an increase in the 3~cm flux density in E7, when $\mdot$ was measured to be the highest by RE19 (Figure~\ref{fig:vla}, right). This possible correlation of the 3~cm emission with $\mdot$ could be attributed to an increase in the jet emission due to an increase in the mass-loss rate ($\mdot$$_{out}$) as $\mdot$ increases. Based on the empirical correlation with the outflow momentum rate \citep{anglada15}, we estimated the variability in the flux of the jet at 3~cm that a variable mass-loss rate would imply. If we assume that $\mdot$$_{out}$ = 10$\%$ $\mdot$, we can calculate the expected flux of the jet. Following \citet{anglada15}, we calculate 4--6~$\mu$Jy, 13--19~$\mu$Jy, and 7--10~$\mu$Jy for the three epochs (assuming a velocity in the jet of 200--300~km/s), which corresponds to changes of $\sim7$--15 and $\sim3$--12~$\mu$Jy between E6 to E7 and E7 to E8. We see changes in GM~Aur of $\sim15$ and $\sim$3~$\mu$Jy. Therefore, the variability in flux could be explained by the variability in the jet. The predicted fluxes from \citet{anglada15} are lower than we measure, but given the uncertainties in the correlation, the likely inclusion of some photoionized disk emission, and the small contribution from dust emission \citep[11 $\mu$Jy,][]{macias16}, the numbers are roughly consistent.  

One issue with this interpretation is the time delay between changes in $\mdot$ and the jet.  This time delay is unknown but is not expected to be instantaneous. However, given our time sampling, we cannot rule out that the $\mdot$ of GM Aur increased over several days before E7 but after E6. Another issue is that radio observations of disks can also trace photoionized gas; high-energy stellar radiation can photoionize the disk, resulting in a wind \citep{alexander14}. \citet{macias16} separated the dust and free-free emission of GM~Aur, showing that the detected free-free emission is produced in two different environments: a photoionized disk and an ionized jet perpendicular to it. Here we do not have the resolution to separate the jet from the photoionized disk.  If the increase in the 3~cm flux is actually due to an increase in the photoionized disk emission, this tentative correlation with $\mdot$ would imply that most of the EUV radiation, predicted to be the main driver of gas photoionization \citep{pascucci12,macias16}, is produced by the accretion shock \citep{calvet98,herczeg07} rather than by the chromospheric activity \citep[e.g.,][]{alexander04}, given that the X-ray emission (and by extension, the chromospheric EUV emission) did not vary between E6 and E8. On the other hand, this lack of X-ray variability supports the interpretation that the 3~cm variability is produced by variability in the jet. 

\section{Summary and Conclusion}

In this first, simultaneous X-ray-UV-optical-IR-cm study of a TTS, we see that as $\mdot$ increases, a robust increase in the MIR emission occurs along with a possible increase in the 3 cm emission. While the 3~cm flux appears to have varied, the X-ray emission remained constant, supportive of the 3~cm variability originating in the jet and not the photoionized disk. The 3~cm emission morphology may have also changed compared with previous observations, pointing to jet variability.  

We conclude that we are seeing the star-disk-jet connection.  An increase in the surface density in the inner disk where material is being loaded onto the star results in a higher $\mdot$ measured on the star and a higher mass-loss rate. This suggests a linked origin, presumably the stellar magnetic field, which can both channel material onto the star as well as eject it in collimated jets along twisted field lines. Future coordinated multiwavelength work is called for to study the connection between mass loading onto the star from the inner disk and subsequent mass ejection via the jet.

 \acknowledgments{ We thank the reviewer for a constructive report.
This work was supported by {\it HST} GO-14048 and GO-15165, SAO GO8-19016A, and the Sloan Foundation. 
}

\end{document}